\pdfoutput=1
\documentclass[journal,twocolumn,10pt]{IEEEtran}

\usepackage[T1]{fontenc}
\usepackage{amsmath,amssymb}
\usepackage{physics}
\usepackage{xspace}
\usepackage{graphicx}
\usepackage[caption=false,font=footnotesize]{subfig}
\usepackage{booktabs}
\usepackage{array}
\usepackage{enumitem}
\usepackage{cite}
\usepackage[hidelinks]{hyperref}

\usepackage[acronym,nogroupskip,nonumberlist,nopostdot]{glossaries}
\glsdisablehyper

\let\originalleft\left
\let\originalright\right
\renewcommand{\left}{\mathopen{}\mathclose\bgroup\originalleft}
\renewcommand{\right}{\aftergroup\egroup\originalright}

\newcommand{\brackets}[1]{{\left[#1\right]}\xspace}

\newcommand{\ydl}{\vy_\dl}
\newcommand{\yul}{\vy_\ul}
\newcommand{\servingbeams}{\qty{\qty(\vf_k^\star,\vw_k^\star)}_{k=1}^K}
\newcommand{\probingbeams}{\qty{\qty(\vf_m, \vw_m)}_{m=1}^M}
\newcommand{\userinfo}{\qty{\qty(\ydl^{(k)}, \yul^{(k)})}_{k=1}^{K}}

\newcommand{\hdl}{\vh_\dl}
\newcommand{\hul}{\vh_\ul}

\newcommand{\useremb}{\mE_\mathrm{U}}
\newcommand{\siemb}{\mE_\si}

\newcommand{\Duser}{D_\mathrm{U}}
\newcommand{\Dsi}{D_\si}

\newcommand{\Rne}{\bar{R}_{\mathrm{eff}}}

\renewcommand{\real}{{\mathbb{R}}}
\renewcommand{\imaginary}{{\mathbb{I}}}
\newcommand{\complex}{{\mathbb{C}}}

\newcommand{\trans}{^{\mathsf{T}}}
\newcommand{\ctrans}{^{{*}}}

\newcommand{\logtwo}[1]{{\mathrm{log}_{2}\qty(#1)}}

\newcommand{\normtwo}[1]{\norm{#1}_{2}}
\newcommand{\normfro}[1]{\norm{#1}_{\mathrm{F}}}
\newcommand{\normmax}[1]{\norm{#1}_{\max}}

\newcommand{\distcgauss}[1]{{\mathcal{N}_{\complex}\qty(#1)}}

\DeclareMathOperator*{\expect}{\mathbb{E}}

\newcommand{\diag}[1]{{\mathrm{diag}\qty(#1)}}

\newcommand{\subjectto}{{\mathrm{s.t.~}}}

\newcommand{\Pnoise}{\sigma^2}

\newcommand{\snr}{{\mathsf{SNR}}}
\newcommand{\sinr}{{\mathsf{SINR}}}
\newcommand{\inr}{{\mathsf{INR}}}

\newcommand{\los}{\mathrm{LOS}}
\newcommand{\nlos}{\mathrm{NLOS}}

\newcommand{\Nt}{{N_\mathrm{t}}}
\newcommand{\Nr}{{N_\mathrm{r}}}

\newcommand{\dl}{\mathrm{DL}}
\newcommand{\ul}{\mathrm{UL}}
\newcommand{\si}{\mathrm{SI}}

\def\vb{{\mathbf{b}}}

\def\vf{{\mathbf{f}}}

\def\vh{{\mathbf{h}}}

\def\vn{{\mathbf{n}}}

\def\vw{{\mathbf{w}}}
\def\vx{{\mathbf{x}}}
\def\vy{{\mathbf{y}}}
\def\vz{{\mathbf{z}}}

\def\vzero{{\mathbf{0}}}

\def\mC{{\mathbf{C}}}

\def\mE{{\mathbf{E}}}
\def\mF{{\mathbf{F}}}

\def\mH{{\mathbf{H}}}
\def\mI{{\mathbf{I}}}

\def\mN{{\mathbf{N}}}

\def\mW{{\mathbf{W}}}
\def\mX{{\mathbf{X}}}
\def\mY{{\mathbf{Y}}}

\def\sP{{\mathcal{P}}}

\def\sS{{\mathcal{S}}}

\newacronym{snr}{SNR}{signal-to-noise ratio}
\newacronym{sinr}{SINR}{signal-to-interference-plus-noise ratio}
\newacronym{inr}{INR}{interference-to-noise ratio}
\newacronym{isac}{ISAC}{integrated sensing and communication}
\newacronym{mrt}{MRT}{maximum ratio transmission}
\newcommand{\MRT}{\gls{mrt}\xspace}

\newacronym{rf}{RF}{radio frequency}
\newacronym{los}{LOS}{line-of-sight}
\newacronym{nlos}{NLOS}{non-line-of-sight}

\newacronym[plural=BSs]{bs}{BS}{base station}
\newacronym{ul}{UL}{uplink}
\newacronym{dl}{DL}{downlink}

\newacronym{mimo}{MIMO}{multiple-input multiple-output}
\newacronym{upa}{UPA}{uniform planar array}

\newacronym{fd}{full-duplex}{in-band full-duplex}
\newacronym{si}{SI}{self-interference}
\newacronym[\glslongpluralkey={sum spectral efficiencies}]{sse}{SSE}{sum spectral efficiency}
\newacronym{lstm}{LSTM}{long short-term memory}

\newcommand{\UPAs}{\glspl{upa}\xspace}

\newcommand{\MIMO}{\gls{mimo}\xspace}
\newcommand{\RF}{\gls{rf}\xspace}
\newcommand{\FD}{\acrshort{fd}\xspace}
\newcommand{\DL}{\acrlong{dl}\xspace}
\newcommand{\UL}{\acrlong{ul}\xspace}
\newcommand{\SI}{\acrlong{si}\xspace}
\newcommand{\BS}{\acrlong{bs}\xspace}
\newcommand{\BSs}{\acrlong{bs}s\xspace}
\newcommand{\SNR}{\gls{snr}\xspace}
\newcommand{\SNRs}{\glspl{snr}\xspace}
\newcommand{\INR}{\gls{inr}\xspace}
\newcommand{\SINR}{\gls{sinr}\xspace}
\newcommand{\SINRs}{\glspl{sinr}\xspace}
\newcommand{\LOS}{\gls{los}\xspace}
\newcommand{\NLOS}{\gls{nlos}\xspace}
\newcommand{\SSE}{\gls{sse}\xspace}

\newcommand{\secref}[1]{Section~\ref{#1}}
\newcommand{\tabref}[1]{Table~\ref{#1}}
\newcommand{\figref}[1]{\figurename~\ref{#1}}

\graphicspath{{./figs/}}

\begin{document}

\title{Site-Specific Beamforming for Full-Duplex Massive MIMO Systems via Implicit Channel Estimation}

\author{Samuel H.~Li and Ian P.~Roberts
\thanks{The authors are with the Wireless Lab, Department of Electrical and Computer Engineering, University of California, Los Angeles (UCLA), Los Angeles, CA USA. Email: \{samuel.li, ianroberts\}@ucla.edu.}
}%

\maketitle

\begin{abstract}
Beamforming has proven to be a promising enabler of full-duplex massive MIMO base stations, but existing beam designs often require knowledge of the self-interference channel matrix $\mH$.
Estimating this high-dimensional channel is costly in practice, however, since it requires a prohibitive number of measurements, especially in fast-fading conditions.
In this work, we overcome this dilemma by designing full-duplex beams using \textit{implicit} channel knowledge gathered from a relatively small number of measurements across $\mH$. 
This is accomplished through site-specific training of a transformer-based deep learning model that learns to efficiently probe portions of $\mH$ most relevant to the particular users being served by exploiting the underlying structure of the surrounding environment. 
The deep learning model then uses these probing measurements to design transmit and receive beams that couple low self-interference while delivering high gain to serve a pair of downlink and uplink users.
For favorable multi-user scaling, a single set of probing measurements can be used by the model to serve several users throughout the coherence time of $\mH$ by leveraging correlations across those users' channels. 
Simulation results using ray-tracing demonstrate that our proposed approach outperforms competitive baselines, as well as the best possible performance with explicit channel estimation, across a range of scenarios, especially with many antennas.%
\end{abstract}%


\glsresetall

\section{Introduction} \label{sec:introduction}

Massive \MIMO wireless transceivers are a cornerstone of 5G communication systems and are expected to be a mainstay in future 6G systems~\cite{andrews_2024_6GTakes,nokia_2025_Transforming6G}.
To meet the demands of emerging applications, there has been recent interest in upgrading massive \MIMO \BSs with in-band \FD capability---the ability to transmit and receive at the same time and same frequency~\cite{smida_2023_FullDuplexWireless}.
This has the potential to increase throughput, extend coverage, reduce latency, and enable monostatic \gls{isac}~\cite{smida_2024_InBandFullDuplexa,nuria_2024_isac},
but is challenged by the manifestation of \textit{self-interference} that couples between the \BS's transmitter and receiver~\cite{kolodziej_2019_Inbandfullduplex}. 
More specifically, self-interference manifests between each \textit{pair} of its transmit and receive antennas, the nature of which depends on near-field coupling as well as reflections off the environment \cite{everett_2014_PassiveSelfInterference,roberts_2024_RealWorld}.
Enabling full-duplex operation requires some method to sufficiently cancel this self-interference~\cite{smida_2024_InBandFullDuplexa,kolodziej_2019_Inbandfullduplex}.

Prior work has shown that, if one has knowledge of all these coupling paths---i.e., the self-interference channel matrix $\mH$---then beamforming can be used to isolate the transmitter and receiver from one another by forming spatial nulls around the channel~\cite{everett_2016_SoftNullManyAntenna,cummings_2020_ApertureLevelSimultaneous,roberts_2021_MillimeterWaveFull}.
Obtaining this self-interference channel state information (CSI) is particularly challenging in practice, however, since the number of required measurements can be prohibitively high, eroding the potential gains offered by \FD operation. 
Given $\Nt$ transmit antennas and $\Nr$ receive antennas at the \BS---each often on the order of tens, hundreds, or even thousands---the number of measurements generally needed to estimate $\mH$ is $\mathcal{O}(\Nt\Nr)$, as the majority of massive \MIMO systems employ analog or hybrid beamforming architectures~\cite{bjornson_2024_6GMIMO,molisch_2017_HybridBeamforming}.
Since reflections off surrounding objects can influence the self-interference channel, estimation of $\mH$ must be repeated at a rate commensurate with the dynamics of the environment, further increasing measurement overhead.
This dilemma marks the focus of this paper: 
How can beams be designed to cancel self-interference and enable \FD massive \MIMO systems without incurring the prohibitive measurement overhead associated with explicitly estimating $\mH$?

\subsection{Related Work}
As mentioned, most existing full-duplex beamforming designs assume (often perfect) knowledge of the self-interference channel.
In \cite{roberts_2021_Hybridbeamforming,koc_2022_IntelligentNonOrthogonal,lopez-valcarce_2022_FullDuplexmmWave,roberts_2023_LoneSTARAnalog,he_2023_FullDuplexCommunication,hernangomez_2026_CISSIR},
for instance, the channel matrix $\mH$ is a required input into optimization problems that solve for beams that maximize spectral efficiency or minimize self-interference.
Other schemes use deep learning methods such as convolutional neural networks \cite{huang_2021_LearningBasedHybrid} and deep-unfolding \cite{bilbao_2024_DeepUnfoldingPowered} to design beams for \FD, yet also require explicit knowledge of $\mH$. 
The overhead associated with estimating $\mH$ and its implications on end performance have been largely overlooked in the vast majority of prior work, however, as they typically report \gls{sinr} or raw spectral efficiency but do not account for the resources spent to estimate $\mH$.

To our knowledge, the only existing designs that do not rely on \textit{explicit} estimation of $\mH$ are \cite{roberts_2022_SteerBeam,roberts_2023_SteerRobust,kong_2024_Activebeam}.
In \cite{roberts_2022_SteerBeam,roberts_2023_SteerRobust}, this was accomplished through a heuristic search of beam candidates, measuring the self-interference coupled by each candidate in order to find those that offer best performance.
This search typically requires 50--500 measurements and needs to be repeated anytime a new user pair is served or when the self-interference channel changes.
A fundamental shortcoming of this approach is that it does not exploit all the spatial degrees of freedom offered by massive \MIMO,
since it searches over predefined beam candidates rather than the entire beam space.
In \cite{kong_2024_Activebeam}, explicit estimation of $\mH$ was circumvented by using a \gls{lstm} architecture to design a sequence of \textit{probing} beams to collect measurements of self-interference that are then used to design actual serving beams.
Like \cite{roberts_2022_SteerBeam,roberts_2023_SteerRobust}, the scheme in \cite{kong_2024_Activebeam} must be repeated for each user pair and thus exhibits poor multi-user scaling.
An additional drawback of \cite{kong_2024_Activebeam} is that it couples measurement of $\mH$ with measurement of the users' channels, which necessitates coordination and feedback from users and thus impedes its practical adoption.

It is worth noting that this problem of high-dimensional channel estimation also appears in \textit{beam management}, where beams need to be steered between a massive \MIMO \BS and a user.
To do so effectively, the two need some knowledge of the downlink/uplink channel, but estimating it directly would similarly incur excessive measurement overhead.
This has led to codebook-based beam management becoming the \textit{de facto} standard in wireless networks such as 5G~\cite{3gpp_5GNR_Phy},
where a set of beams are swept to identify the one which maximizes \gls{snr}.
Within this beam-sweeping framework, researchers have proposed sparse probing techniques \cite{3gpp_5GNR_Phy,heng_2024_Sitespecificbeam,xue_2024_SurveyBeam,heng_2024_GridFreeMIMO,alrabeiah_2022_NeuralNetworks,dreifuerst_2024_MachineLearning,kwak_2024_SiteSpecific},
which leverage deep learning to design small sets of probing beams tailored to particular environments in place of predefined beam codebooks;
we take a similar approach in this paper.
In other works~\cite{va_2018_InverseMultipath, Demirhan_2022_RadarAided, jiang_2022_ComputerVision, yang_2025_RadioMapBased},
machine learning has been used to infer the optimal beam from side information such as GPS, camera, radar, lidar, or lower-frequency channel data. 
Also of note are compressive sensing techniques to estimate high-dimensional channels or identify angles-of-departure/arrival~\cite{alkhateeb_2014_ChannelEstimation,lee_2016_ChannelEstimation}.
The primary drawbacks of such approaches include their reliance on channel sparsity and their sensitivity to noise in practice;
they can also require a prohibitively large number of measurements for practical deployments~\cite{alkhateeb_2015_Compressedsensing,sohrabi_2022_ActiveSensing}.

\subsection{Contributions}

In this work, we introduce a novel approach to designing the transmit and receive beams of a full-duplex massive \MIMO \BS that circumvents explicit estimation of the self-interference channel $\mH$.
We employ a deep learning model to first design a set of \textit{probing} beams that are swept by the \BS to collect $M$ measurements of self-interference, providing it with \textit{implicit} knowledge of the underlying channel $\mH$;
these measurements are then processed by the model to synthesize transmit and receive beams which serve pairs of downlink and uplink users in a full-duplex fashion.
Rather than repeat probing measurements for each user pair, we architect our solution such that a single set of $M$ measurements can be reused across multiple user pairs for favorable scaling.%
\footnote{This multi-user aspect is a key functional difference between the present paper and our conference paper~\cite{li_2025_SiteSpecificBeam}. As a result of this, the deep learning implementation herein is substantially more sophisticated than that in~\cite{li_2025_SiteSpecificBeam}.}

Our deep learning solution consists of multiple transformer-based modules that are trained end-to-end in a site-specific fashion, tailoring it to the particular environment in which the \BS is deployed.
This allows the model to learn to craft probing beams that exploit the underlying structure of the environment in order to efficiently probe the portions of $\mH$ that are most relevant to the particular group of users to be served.
The implicit channel knowledge contained in these measurements can then be interpreted by the model to synthesize serving beams to cancel self-interference while delivering high \gls{snr} to the users.

Evaluation via ray-tracing simulations demonstrates that our model synthesizes near-optimal serving beams using only a fraction of the measurements required to explicitly estimate the self-interference channel $\mH$.
When accounting for measurement overhead, our proposed approach consistently outperforms competitive baselines across a wide range of operational regimes.
In a representative setting with eight user pairs, for example, our solution achieves over 80\% of the full-duplex capacity with only $M=16$ probing measurements, whereas a competitive baseline using sparse estimation falls below 70\%, and the theoretical upper bound with explicit channel matrix estimation is capped below 64\%.
Furthermore, our approach scales gracefully to large antenna arrays, translating the additional spatial degrees of freedom into net performance gains with no additional probing measurements and only minor increases in computational complexity.
Finally, our model exhibits inherent robustness to deployment mismatches by maintaining strong performance in the presence of blockage variations and physical misalignment of the antenna array.

\begin{figure}[!t]
  \centering
  \includegraphics{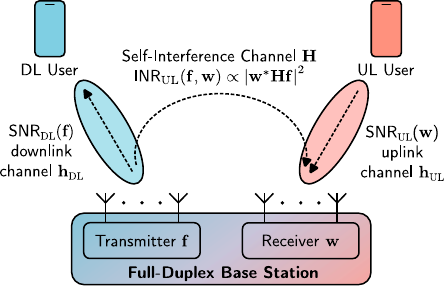}
  \caption{An in-band \FD \BS transmits to a \DL user while receiving from an \UL user across the same frequency band.
  This work aims to optimally design $\vf$ and $\vw$ without \textit{explicitly} estimating $\mH$.}
  \label{fig:system-model}
\end{figure}

\subsection{Notation}
The following notation is used throughout this paper:
standard Roman font $x$ is used to denote scalars;
bold lowercase $\vx$ is used to denote vectors;
bold uppercase $\mX$ is used to denote matrices;
$[\vx]_i$ denotes the $i$-th element of a vector $\vx$;
$\real$ and $\complex$ denote the sets of real and complex numbers;
$\abs{\cdot}$, $\norm{\cdot}_1$, $\normtwo{\cdot}$, $\normmax{\cdot}$, and $\normfro{\cdot}$ represent element-wise absolute value, $\ell_1$-norm, $\ell_2$-norm,  max-norm, and Frobenius norm, respectively.

\section{System Model}
\label{sec:system-model}

As illustrated in \figref{fig:system-model}, we consider an in-band \FD \BS equipped with separate transmit and receive arrays, serving single-antenna downlink and uplink users.
The \BS's transmit and receive arrays consist of $\Nt$ and $\Nr$ antenna elements, respectively, each driven by a single \RF chain and equipped with analog beamforming.
The \BS transmits to the downlink user using transmit beam $\vf \in \complex^{\Nt \times 1}$, while simultaneously receiving from the uplink user with receive beam $\vw \in \complex^{\Nr \times 1}$.
The resulting downlink and uplink \SNRs are given by
\begin{align}
  \label{eq:snr-dl}
  \snr_\dl \qty(\vf) & = \frac{P_{\dl} \, \abs{\hdl^{*}\vf}^{2}}{\Nt \, \Pnoise_{\dl}},
  \\
  \label{eq:snr-ul}
  \snr_\ul \qty(\vw) & = \frac{P_{\ul} \, \abs{\vw^{*}\hul}^{2}}{\normtwo{\vw}^{2} \, \Pnoise_{\ul}}.
\end{align}
Here, $\hdl \in \complex^{\Nt \times 1}$ and $\hul \in \complex^{\Nr \times 1}$ are the downlink and uplink channel vectors;
$P_\dl$ and $P_\ul$ are the transmit powers of the \BS and of the uplink user;
$\Pnoise_\ul$ and $\Pnoise_\dl$ are the noise variances at the \BS and at the downlink user, respectively.
The divisions by $\Nt$ in \eqref{eq:snr-dl} and by $\normtwo{\vw}^2$ in \eqref{eq:snr-ul} account for power splitting across the transmit antennas and noise combining at the receive array, respectively.

We assume analog beamforming is implemented via a network of analog phase shifters and variable attenuators, and this imposes a per-antenna power constraint on the transmit beam $\vf$ as
\begin{align}
  \label{eq:power-constriant}
  \abs{[\vf]_i} & \leq 1, \quad  i = 1, \ldots, \Nt.
\end{align}

By transmitting and receiving in a full-duplex fashion, the \BS inflicts interference upon itself across the \MIMO channel matrix $\mH \in \complex^{\Nr\times\Nt}$.
We model $\mH$ as the sum of separate \LOS and \NLOS components by \cite{duarte_2012_ExperimentDrivenCharacterization,roberts_2024_RealWorld}
\begin{equation}
  \label{eq:si-channel}
  \mH = \sqrt{\frac{\kappa}{\kappa+1}} \, \mH_\los + \sqrt{\frac{1}{\kappa+1}} \, \mH_\nlos,
\end{equation}
where $\kappa$ is the Rician factor that controls the relative strength of the \LOS and \NLOS components.
The \LOS component $\mH_\los$ captures near-field coupling between each pair of antennas in the arrays and is largely determined by the array geometry at the \BS; it can therefore be assumed quasi-static \cite{roberts_2022_BeamformedSelfInterference}.
The \NLOS component $\mH_\nlos$, on the other hand, captures multipath reflections from the surrounding environment and is assumed to vary more rapidly with time.

For a given transmit beam $\vf$ and receive beam $\vw$, the strength of \SI incurred at the \BS is proportional to $|\vw\ctrans\mH\vf|^2$, with the uplink \INR putting this relative to noise as
\begin{equation}
  \label{eq:inr}
  \inr_\ul \qty(\vf, \vw) = \frac{P_\dl \, \abs{\vw\ctrans \mH
  \vf}^{2}}{\Nt \, \normtwo{\vw}^{2} \, \Pnoise_{\ul}}.
\end{equation}

In this full-duplex setting, the downlink user may also experience cross-link interference from the uplink user's transmission, with the downlink \INR given by
\begin{equation}
  \label{eq:inr-dl}
  \inr_\dl = \frac{P_{\ul} \, \abs{h}^{2}}{\Pnoise_{\dl}},
\end{equation}
where $h \in \complex$ denotes the scalar interference channel between the \UL and \DL users.
Cross-link interference is typically orders of magnitude weaker than that of \SI, due to larger distances between users and lower uplink transmit powers \cite{roberts_2024_RealWorld,sabharwal_2014_InBandFullDuplex},
making \SI our primary focus.

Accounting for the effects of  additive noise and interference, we can formulate the downlink and uplink \SINRs as
\begin{align}
  \label{eq:sinr-dl}
  \sinr_\dl \qty(\vf) &= \frac{\snr_\dl \qty(\vf)}{1 + \inr_\dl}, \\ 
  \label{eq:sinr-ul}
  \sinr_\ul \qty(\vf, \vw) &= \frac{\snr_\ul \qty(\vw)}{1 + \inr_\ul \qty(\vf, \vw)}.
\end{align}
The \SINRs dictate the achievable spectral efficiency on each link, which can be expressed as
\begin{align}
  \label{eq:se-dl}
  R_\dl(\vf)     & = \logtwo{1+\sinr_\dl \qty(\vf)},
  \\
  \label{eq:se-ul}
  R_\ul(\vf,\vw) & = \logtwo{1+\sinr_\ul \qty(\vf, \vw)}.
\end{align}
Overall system performance can then be measured by the \SSE
\begin{equation}
  \label{eq:sse}
  R(\vf, \vw) = R_\dl(\vf) + R_\ul(\vf, \vw),
\end{equation}
which jointly captures the effects of downlink \SNR, uplink \SNR, self-interference, and cross-link interference as a function of the \BS's transmit and receive beams $(\vf,\vw)$.%

\begin{figure*}[t]
  \centering
  \includegraphics[width=\linewidth,height=\textheight,keepaspectratio]{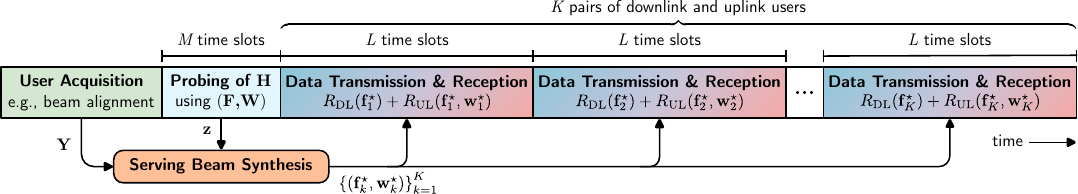}%
  \caption{Timeline of the envisioned use of our proposed scheme,
  with one time slot defined as the time to collect a single probing measurement across $\mH$.}
  \label{fig:timeline}
\end{figure*}

\section{Problem Statement} \label{sec:problem-formulation}

The goal of this work is to create a scheme which designs the transmit beam $\vf$ and receive beam $\vw$ of the full-duplex \BS to maximize the \SSE when serving pairs of downlink and uplink users over time.
Within this context, we consider the case where the \SI channel $\mH$ varies with time according to a block-fading model.
More specifically, the coherence time of $\mH$ is defined such that $K$ user pairs can be served, each for $L$ time slots, before the channel changes.
We term this a \textit{coherent user group}.
If the \BS collects $M$ measurements of self-interference to estimate $\mH$ before serving these $K$ user pairs, then the mean \textit{effective \acrlong{sse}} among these users is equal to
\begin{align}\label{eq:eff-sse}
    \frac{KL}{M+KL} \pqty{\frac{1}{K} \sum_{k=1}^K R_k(\vf_k^\star,\vw_k^\star)},
\end{align}
where $(\vf^\star_k,\vw^\star_k)$ are the transmit and receive beams used to serve the $k$-th user pair and $R_k(\vf^\star_k,\vw^\star_k)$ is the resulting \SSE delivered to that user pair.
The principal objective of this work is to design near-optimal serving beams $\servingbeams$ in order to maximize the \SSE to those $K$ user pairs.

From \eqref{eq:eff-sse}, we can observe the core trade-off at the center of this work.
Designing $(\vf_k^\star,\vw_k^\star)$ to maximize spectral efficiency requires knowledge of self-interference,
but if too many measurements $M$ are used to gather this knowledge, the effective \SSE will be degraded~\cite{hassibi_2003_Howmuch}.
This motivates optimizing the $M$ \textit{probing} beam pairs $\probingbeams$ that are used to collect these measurements.
We aim to show that, through such tailored probing, the \BS can achieve performance comparable to that obtained via explicit estimation of $\mH$, while using far fewer measurements, i.e., $M \ll \Nt\Nr$.
As shall be seen, the key to fulfilling this aim is our use of deep learning to \textit{jointly} design the $M$ probing beam pairs and the final $K$ serving beam pairs.

\section{Site-Specific Beamforming via Implicit Self-Interference Channel Estimation} \label{sec:beam-learning}

\begin{figure*}[!t]
  \centering
  \includegraphics[width=\textwidth]{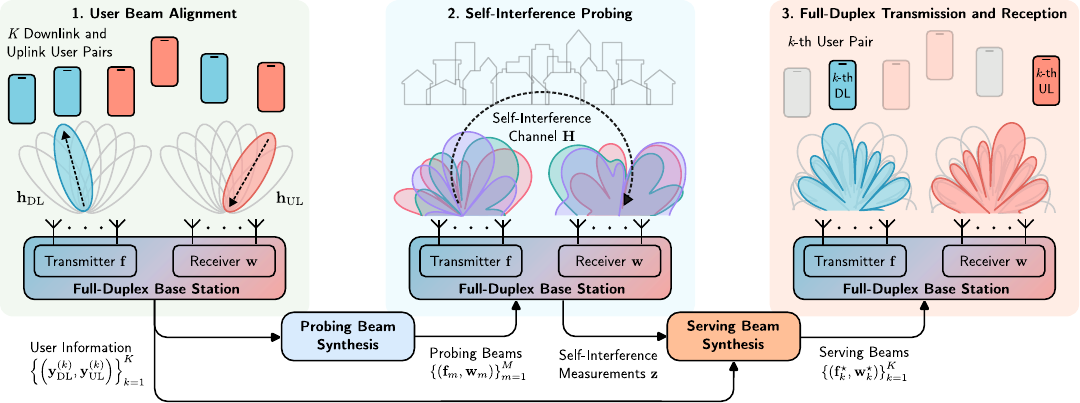}
  \caption{Our proposed approach consists of three stages. First, beam alignment is used to obtain knowledge of the downlink and uplink users' channels. Then, using this user channel information, a sequence of $M$ probing beam pairs are designed to collect $M$ measurements across the self-interference channel~$\mH$. Finally, these measurements are used along with the user information to synthesize transmit and receive beams to serve user pairs in a full-duplex fashion.}
  \label{fig:stages}
\end{figure*}

In this section, we detail our proposed full-duplex beamforming design, which aims to design the $M$ probing beam pairs $\left\{(\vf_m,\vw_m)\right\}_{m=1}^M$ and the final $K$ serving beam pairs $\left\{(\vf^\star_k,\vw^\star_k)\right\}_{k=1}^K$.
As shown in \figref{fig:timeline} and \figref{fig:stages}, our envisioned scheme consists of three main components:
\begin{enumerate}
  \item Conventional beam alignment (or a similar procedure) to obtain some knowledge of the users' channels.
  \item Measurements of self-interference taken with $M$ probing beam pairs to obtain \textit{implicit} channel knowledge of $\mH$.
  \item Synthesis of $K$ transmit and receive beams to serve $K$ pairs of downlink and uplink users in a full-duplex fashion.
\end{enumerate}
We overview each of these components, along with our final design problem, in the following subsections.
Then, in the next section, we elaborate on the deep learning implementation used to realize this envisioned solution.

\subsection{User Acquisition}
Like in conventional half-duplex systems, the full-duplex \BS requires knowledge of the users' downlink and uplink channels in order to serve them with high gain.
Our proposed framework begins by assuming that existing mechanisms are used to obtain this information, such as beam alignment in 5G networks~\cite{xue_2024_SurveyBeam}.
In general, this channel knowledge can either be explicit estimates of $\vh_\dl$ and $\vh_\ul$ for each user pair or implicit knowledge of some form, such as raw beam alignment measurements or angles-of-departure/arrival.
We denote this user channel information across all $K$ user pairs by $\{(\ydl^{(k)},\yul^{(k)})\}_{k=1}^K$.
Naturally, gathering this channel knowledge would need to be repeated commensurate with the coherence time of the users' channels.
We remark that, by not mandating a particular strategy for obtaining this user channel knowledge, our proposed scheme seeks to maintain backward compatibility with existing standards and facilitate its adoption in real-world networks.

\subsection{Self-Interference Probing}
The next stage of our proposed solution is to collect $M$ measurements across the self-interference channel $\mH$.
To do so, we design $M$ probing beam pairs $\left\{(\vf_m,\vw_m)\right\}_{m=1}^M$, where $(\vf_m,\vw_m)$ are the transmit and receive beams used by the \BS to collect the $m$-th measurement of self-interference, %
which can be expressed as
\begin{align}
  \label{eq:si-probing-single}
  z_m = \sqrt{\frac{P_\dl}{\Nt}} \, \vw_m^* \mH \vf_m + \vw_m^* \vn_m,
\end{align}
where $\vn_m \sim \distcgauss{\vzero, \sigma^2_\ul \mI}$ denotes complex Gaussian noise at the receive array.
All $M$ measurements of self-interference can then be written compactly in vector form as
\begin{align}
  \label{eq:si-probing-vec}
  \vz & = \sqrt{\frac{P_\dl}{\Nt}} \, \diag{\mW\ctrans \mH \mF}
  + \diag{\mW\ctrans \mN}
  \in \complex^{M \times 1},
\end{align}
with $\vz = [z_1 \cdots z_M]\trans$, $\mN = [\vn_1 \cdots \vn_M]$, and the transmit and receive probing beams stacked into the matrices
\begin{align}
  \mF &= \brackets{\vf_1 \ \cdots \ \vf_M} \in \complex^{\Nt \times M}, \\
  \mW &= \brackets{\vw_1 \ \cdots \ \vw_M} \in \complex^{\Nr \times M},
\end{align}
which we refer to as the transmit and receive \textit{probing codebooks}.
Note that these $M$ measurements of self-interference will need to be recollected once $\mH$ changes, 
which we assume is after the $K$ user pairs have been served. %

Core to our approach is its tailoring of the probing codebooks $\mF$ and $\mW$ to the particular $K$ user pairs that the \BS intends to serve.
As detailed in the next section, this is carried out by a deep learning model which performs a mapping $\sP(\cdot)$ from user channel knowledge to probing codebooks, i.e.,
\begin{equation}
  \label{eq:P}
  (\mF,\mW) = \sP(\mY),
\end{equation}
where
\begin{equation}
  \label{eq:Y}
  \mY \triangleq \userinfo.
\end{equation}
Even though the self-interference channel $\mH$ itself does not depend on the particular users being served, the amount of self-interference that is ultimately coupled does,
since the \BS will form its transmit and receive serving beams based on the users' channels.
While explicitly estimating the entire matrix $\mH$ would provide knowledge of the self-interference coupled by \textit{any} possible serving beams $(\vf,\vw)$, we hypothesize that focusing only on the subspace of $\mH$ that is particularly relevant to these specific users will allow the \BS to attain comparable performance with substantially fewer measurements, compared to obtaining full knowledge of $\mH$.
We confirm this to indeed be true in our performance evaluation in \secref{sec:results}.

\subsection{Serving Beam Synthesis}

The final step of our proposed approach lies in generating the transmit and receive beams used by the \BS to serve each of the $K$ user pairs in a full-duplex fashion.
Like the probing beams, the serving beams will be designed using deep learning via a policy $\sS(\cdot)$, which takes as input the self-interference probing measurements $\vz$ and the user channel knowledge $\mY$, i.e.,
\begin{align}
  \label{eq:S}
  \servingbeams = \sS(\vz,\mY).
\end{align}
As mentioned, these serving beams must deliver high \gls{snr} to the users while coupling low self-interference across $\mH$ in order to enable full-duplex operation.
To design such, we will jointly optimize $\sP(\cdot)$ and $\sS(\cdot)$ in order to design probing beams that gather implicit knowledge of $\mH$ that is the {most informative} for synthesizing optimal serving beams.

\subsection{Problem Formulation}

To formalize the design of the probing beam policy $\sP(\cdot)$ and serving beam policy $\sS(\cdot)$ based on our stated goals, we assemble the following optimization problem.
\begin{subequations} \label{eq:problem}
  \begin{align}
    \label{eq:objective}
    \max_{\sP, \sS}
    \quad
    & \expect_{\mH,\mY} \qty[\sum_{k=1}^K \bar{R}_k(\vf^\star_k, \vw^\star_k)]
    \\ \subjectto \
    & (\mF, \mW) = \sP(\mY)
    \label{eq:constr-probegen}
    \\ \quad
    & \servingbeams = \sS(\vz, \mY)
    \label{eq:constr-beamsyn}
    \\ \quad
    &  \vz = \sqrt{\frac{P_\dl}{\Nt}} \, \diag{\mW\ctrans \mH \mF} + \diag{\mW\ctrans \mN}
    \label{eq:constr-probe}
    \\ \quad
    & \normmax{\vf^\star_k} \leq 1, \quad k = 1, \dots, K
    \label{eq:constr-power-beam}
    \\ \quad
    & \normmax{\vf_m} \leq 1, \quad m = 1, \dots, M
    \label{eq:constr-power-probe}
  \end{align}
\end{subequations}
Here, our objective function has been defined as the expected sum of the \textit{normalized} spectral efficiency across user pairs rather than the raw spectral efficiency as in \eqref{eq:eff-sse}.
This normalized \SSE is defined as
\begin{align}
  \bar{R}_k(\vf^\star_k,\vw^\star_k) &= \frac{R_k(\vf^\star_k,\vw^\star_k)}{C_k},
\end{align}
with $C_k$ the interference-free sum capacity of the $k$-th user pair, defined as
\begin{align}
  \label{eq:capacity}
  C_k & = \logtwo{1 + \frac{P_\dl \normtwo{\hdl^{(k)}}^2}{\Pnoise_\dl}} + \logtwo{1 + \frac{P_\ul \normtwo{\hul^{(k)}}^2}{\Pnoise_\ul}}.
\end{align}
This normalization promotes fairness across user pairs, so that those with stronger channels are not excessively favored.
Normalizing by the interference-free capacity places user pairs on a common scale while still preserving the diminishing returns associated with the logarithmic growth of spectral efficiency, encouraging our solution to cater to low-\SNR users, since they stand to benefit proportionally more from gains in \SINR.
Nonetheless, other normalizations or objective functions could be used if preferred.
Note that the expectation of the objective function is over the distributions of $\mH$ and $\mY$.
This will ensure that a single design of $\sP(\cdot)$ and $\sS(\cdot)$ will be useful regardless of the particular self-interference channel $\mH$ and user grouping at any given time.

In problem~\eqref{eq:problem},
line \eqref{eq:constr-probegen} tailors the $M$ probing beams to the group of $K$ user pairs to be served;
line \eqref{eq:constr-probe} represents the $M$ probing measurements of self-interference taken across $\mH$;
line \eqref{eq:constr-beamsyn} synthesizes the $K$ beam pairs used to serve the $K$ user pairs;
and lines \eqref{eq:constr-power-beam}--\eqref{eq:constr-power-probe} enforce per-antenna power constraints on the transmit beams.
Recognize that this problem would be solved for a fixed number of measurements $M$ and number of user pairs $K$, both of which may be manually tuned by network engineers to optimize performance based on operating conditions, as we will see in our performance evaluation in \secref{sec:results}.
In the next section, we aim to solve problem~\eqref{eq:problem} through a transformer-based deep learning implementation.

Note that, by limiting the scope of problem~\eqref{eq:problem} to a single \BS, $\sP(\cdot)$ and $\sS(\cdot)$ can be designed in a \textit{site-specific} fashion, tailoring them to the particular self-interference distribution and user distribution seen by that \BS
---both of which depend on features of the surrounding environment, such as buildings, vehicles, foliage, etc.
In practice, this would be accomplished by training our model using channel data from measurements, a statistical model, a ray-tracing simulator, and/or digital twin; in the performance evaluation in \secref{sec:results}, we use ray-tracing.
Our approach to solving \eqref{eq:problem} is therefore data-driven, optimizing the policies $\sP(\cdot)$ and $\sS(\cdot)$ directly over channel realizations rather than relying on closed-form channel models or analytical approximations.
Although not strictly necessary, this site-specific design minimizes the number of measurements $M$ by restricting both the user and self-interference channels to those observed at a particular \BS, enabling the model to learn and exploit the spatial structure of the local propagation environment.

\section{Deep Learning Implementation} \label{sec:implement}

\begin{figure}[!t]
  \centering
  \includegraphics{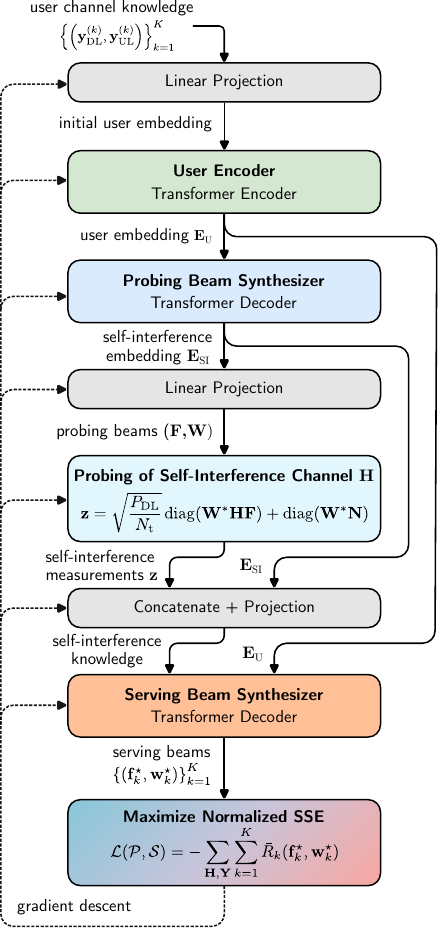}
  \caption{Proposed transformer-based deep learning model to realize our envisioned site-specific beamforming solution. The entire model is trained end-to-end to jointly design the probing and serving beam synthesizers.}
  \label{fig:ml-model}
  \vspace{-0.1cm}
\end{figure}

In this section, we detail the deep learning implementation that we have developed to solve problem \eqref{eq:problem} in a site-specific fashion.
Corresponding to the three components discussed in \secref{sec:beam-learning}, our proposed deep learning framework consists of three modules: the \textit{user encoder}, the \textit{probing beam synthesizer}, and the \textit{serving beam synthesizer}, as illustrated in \figref{fig:ml-model}.
All three modules are transformer-based \cite{vaswani_2017_AttentionAll}, as detailed next.

\subsection{User Encoder} \label{subsec:user-encoder}

The first module is the \textit{user encoder},
which uses a transformer encoder to act as a dedicated pre-processor that maps the user information $\mY$ to a \textit{user embedding} tensor $\useremb$ that is used in subsequent tasks.
This embedding can be thought of as a compact, real-valued representation of the information about the users' channels that is most relevant in designing the probing and final serving beams, with each user pair's embedding having a fixed dimension $\Duser$.

Since each of the $K$ entries of $\mY$ consists of two complex-valued vectors with dimensions equal to the number of transmit and receive antennas, they need to be converted to fixed-length, real-valued inputs $\mY_\real = \{\vy_\real^{(k)}\}_{k=1}^K$ before being ingested by the encoder.
To do so, the real and imaginary components of $\vy_\dl$ and $\vy_\ul$ are concatenated jointly along the array dimension to form a single real-valued vector of length $2(\Nt+\Nr)$, i.e.,
\[
  \vy_\real^{(k)} = \qty[\real\qty{\vy_{\dl}^{(k)}}\trans, \imaginary\qty{\vy_{\dl}^{(k)}}\trans, \real\qty{\vy_{\ul}^{(k)}}\trans, \imaginary\qty{\vy_{\ul}^{(k)}}\trans]\trans.
\]

A linear projection layer subsequently reduces the dimension of each $\vy_\real^{(k)}$ from $2(\Nt+\Nr)$ to $\Duser$ to form the initial user embeddings.
The transformer encoder then iteratively refines $\useremb$ through its self-attention layers, which selectively aggregate channel information across all $K$ user pairs.
The resulting $\useremb$ can thus contain valuable knowledge of not just the users' channels but the correlations across the coherent user group, which can be exploited when designing the probing and serving beams.

Note that, by operating on the fixed-dimension $\useremb$ rather than $\mY_\real$ directly,
the complexity of the transformer is independent of the number of transmit/receive antennas.
Consequently, a single user encoder can accommodate varying antenna counts. 
Lightweight linear projections first map user information from different array sizes to embeddings of fixed size $\Duser$, which are then processed by a common transformer backbone trained jointly across diverse configurations and operating conditions.
Additionally, no positional encoding or causal masking is applied within the self-attention layers,
since all $K$ pairs of user channels are available to the user encoder from the start,
and their ordering does not carry semantic meaning in this context,
so these embeddings should be attended to in parallel and in a permutation-invariant manner.

\subsection{Probing Beam Synthesizer}

After the user encoder, the \textit{probing beam synthesizer} is responsible for realizing the probing policy $\sP(\cdot)$.
It conditions on the user embedding $\useremb$ to generate the \SI probing beam codebooks $(\mF,\mW)$ for the particular coherent user group.
At the core of the probing synthesizer is a transformer decoder, which generates a set of $M$ \SI embeddings $\siemb$, each corresponding to a particular pair of probing beams $(\vf_m,\vw_m)$ to be generated.

Since the \BS has no instantaneous knowledge of the particular $\mH$ before probing, it must rely on its prior understanding of the channel's distribution, which is learned during training as will be discussed shortly.
Alongside the model weights, this understanding is also captured through a learnable embedding table from which the initial $M$ embedding vectors are drawn, thus providing a unique starting point for each of the $M$ probing beam pairs.

The \SI embeddings are subsequently passed through the transformer decoder with alternating self-attention and cross-attention layers.
The cross-attention layer takes $\useremb$ as the key--value memory,
pivoting each \SI embedding to focus on the components of $\mH$ most relevant to these particular users.
The self-attention then interchanges information among $\siemb$, so that probing beams yield \textit{complementary} measurements rather than overlapping ones.
Finally, a projection layer maps the $m$-th \SI embedding to the real and imaginary components of the corresponding probing beams to construct $(\vf_m, \vw_m)$.

Since the output of the final projection layer does not necessarily satisfy the per-antenna power constraint in \eqref{eq:constr-power-probe}, we explicitly normalize each transmit probing beam.
To do so, each transmit probing beam vector is scaled by its maximum entry magnitude, e.g.,
\begin{equation}
  \label{eq:beam-normalization}
  [\vf_m]_{i} = \frac{[\tilde{\vf}_m]_{i}}{\max_{j} | [\tilde{\vf}_m]_{j} |},
\end{equation}
where $\tilde{\vf}_m$ is the $m$-th unnormalized transmit probing beam output by the model and $\vf_m$ is its normalized version that populates the $m$-th column of $\mF$.
We normalize $\mW$ the same way.
We observed that other normalizations, such as simple clipping, can also be used and offer similar performance.
The normalized probing beam codebooks $\mF$ and $\mW$ are then used to collect $M$ measurements of \SI, populating $\vz$ according to \eqref{eq:si-probing-vec}.

In a similar manner to the user encoder, the probing beam pairs in $\mF$ and $\mW$ are generated in parallel:
the model generates the $m$-th beam pair $(\vf_m,\vw_m)$ conditioned only on $\useremb$ and not measurements from ``preceding'' beams, e.g., $(z_1, \dots, z_{m-1})$.
Rather than generate probing beam pairs autoregressively, like in \cite{kong_2024_Activebeam}, this parallel approach was done deliberately to avoid the latency associated with processing and inference between measurements, so that probing measurements can be collected in rapid succession.
Furthermore, because probing beam synthesis depends solely on $\useremb$, all $M$ probing beam pairs can be synthesized in the background as soon as the coherent user group has been scheduled.

\subsection{Serving Beam Synthesizer}

The final module of our proposed solution is the \textit{serving beam synthesizer},
which realizes the beam synthesis policy $\sS(\cdot)$.
It takes as input the user embedding $\useremb$, the \SI embedding $\siemb$, and the probing measurements $\vz$,
and generates the final serving beams $\servingbeams$ for the coherent user group.
It is implemented as a transformer decoder similar to the probing beam synthesizer, but with subtle differences to its input and output structure.
Here, the user embedding $\useremb$ is supplied directly as the initial embedding into the transformer decoder.
To condition on the collected probing measurements, the real and imaginary components of $\vz$ are concatenated with $\siemb$ and projected as the key--value memory to the cross-attention layers, which condition the beam embeddings on the implicit knowledge of $\mH$ to steer the final serving beams away from \SI.
In this sense, our approach can be viewed as measurement-conditioned beam synthesis, with the model learning task-oriented latent representations of the self-interference channel, rather than explicitly estimating its physical parameters.
Finally, a projection layer maps the final beam embeddings to the real and imaginary components of the serving beam pairs, which are then normalized according to \eqref{eq:beam-normalization} to produce the final beams $\servingbeams$.
Like the probing beams, the serving beams are generated in a non-autoregressive manner, in parallel, and without positional encoding.
We would like to point out that, if the coherence time of $\mH$ is sufficiently long relative to $KL$, then the proposed approach need not follow the nominal timeline of \figref{fig:timeline} exactly.
Rather, the base station can collect the probing measurements $\vz$ somewhat in advance, allowing the serving beam synthesizer to execute in the background while other transmissions are ongoing.
This would ensure that the serving beams are computed by the time the users need to be served, in case inference latency is otherwise a limiting factor.

\subsection{Site-Specific Training \& Architectural Intuition}

Our entire deep learning model is trained end-to-end in an unsupervised manner to maximize the sum normalized spectral efficiency via the loss function
\begin{equation}
    \label{eq:loss}
    \mathcal{L}(\sP, \sS) = - \sum_{\mH, \mY} \sum_{k=1}^K \bar{R}_k(\vf^\star_k, \vw^\star_k).
\end{equation}
Here, batching is done over random user groupings and \SI channel realizations to approximate the expectation in \eqref{eq:objective}.
Because our problem is non-convex, gradient descent is not guaranteed to find a global optimum.
To promote stable convergence and avoid local optima, we use mini-batch stochastic training with the AdamW optimizer~\cite{loshchilov_2019_DecoupledWeight} and learning rate scheduling.

We would like to remark that the proposed transformer-based architecture allows a single model to generalize across a variety of configurations and operating conditions, in addition to antenna count, as described in \secref{subsec:user-encoder}.
For instance, since $K$ and $M$ affect the sequence \textit{lengths}, e.g., the number of vectors within their embeddings, rather than the \textit{dimensions} of these vectors, a single trained model can be used for various coherent user group sizes and probing budgets.
We take advantage of this via a two-stage training process where a shared model is first {pretrained} broadly across diverse configurations of $(K,M,\Nt,\Nr)$,
then {fine-tuned} to the specific deployment setting at a fraction of the cost of training from scratch.
This broad pretraining provides a stable initialization that further mitigates the risk of suboptimal convergence.
During fine-tuning, the model is trained on a particular configuration of $(K,M,\Nt,\Nr)$, tailoring it to the target operational setting; we find this fine-tuning generally improves performance by only a few percent.

\section{Simulation Setup and Baselines} \label{sec:simulation}

\begin{figure}[t]
  \centering
  \includegraphics[width=\linewidth]{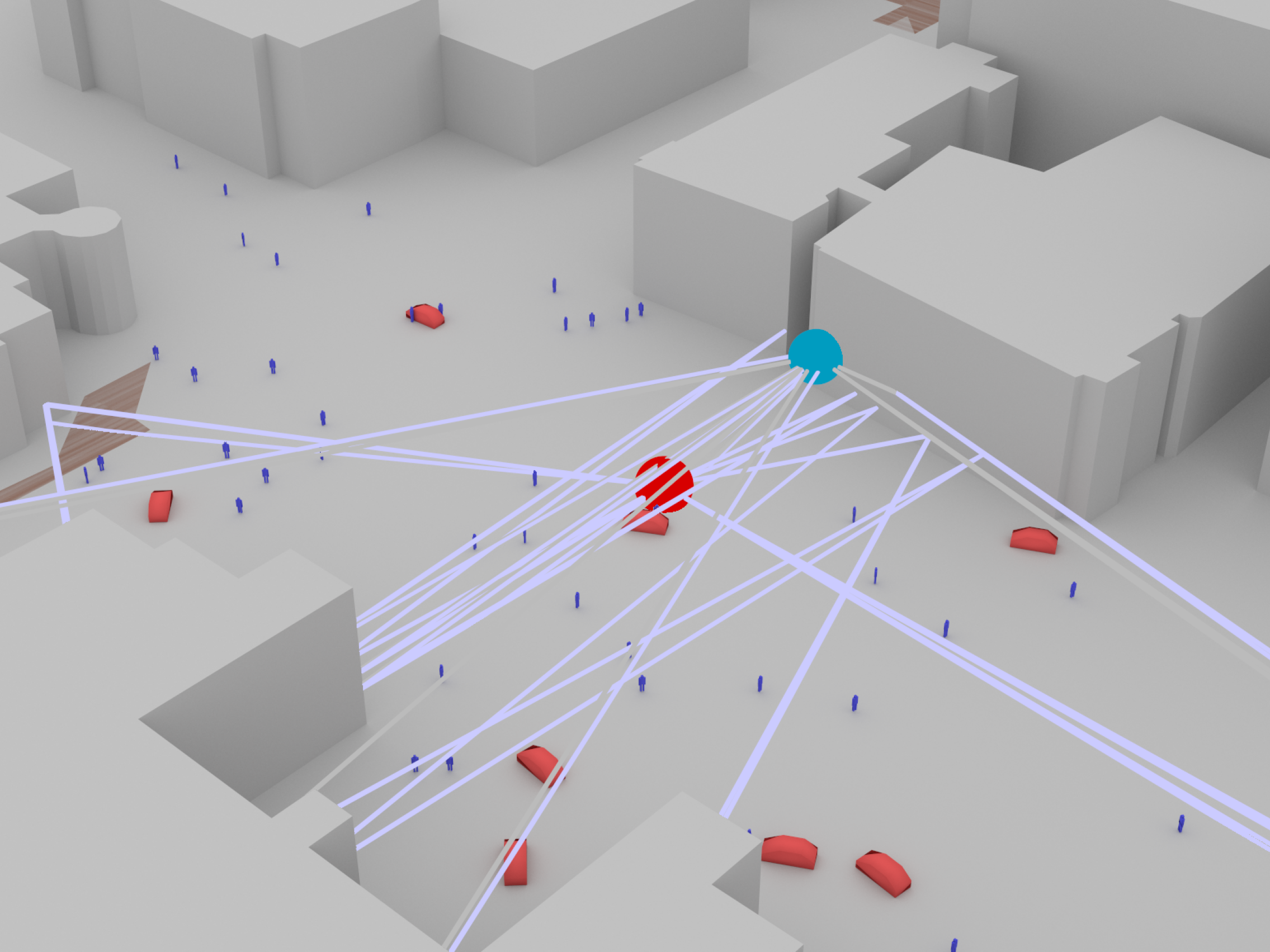}
  \caption{
    A sample scene realization based on a reconstructed 3D model of Westwood Plaza on the UCLA campus, derived from OpenStreetMap~\cite{haklay_2008_OpenStreetMapUserGenerated} data.
    The cyan and red spheres indicate the locations of the \BS atop a building and a downlink user in the courtyard, respectively.
    The blue figures represent pedestrians and the red objects represent vehicles,
    both randomly placed in the courtyard.
    The purple lines indicate representative ray-traced propagation paths between the \BS and the user equipment.
    For readability, the rest of the downlink and uplink users are omitted from visualization but remain present in the simulation.
    }
  \label{fig:scenario}
\end{figure}

To evaluate the performance of our proposed approach, we simulate a \FD massive \MIMO \BS operating at a carrier frequency of 28~GHz.
The \BS employs separate transmit and receive antenna arrays equipped with analog beamforming.
The arrays are placed on the same plane facing the center of the user distribution, with their centers separated horizontally by 10 wavelengths.
Unless otherwise specified, we use 4$\times$4 half-wavelength \UPAs for both the transmitter and receiver, yielding $\Nt=\Nr=16$ elements; we investigate performance with larger arrays shortly.

In an effort to provide realistic performance insights, we perform ray-traced channel modeling using NVIDIA's Sionna Ray Tracing library~\cite{hoydis_2022_Sionna}.
As shown in \figref{fig:scenario}, we conduct simulations in a reconstructed 3D scene of Westwood Plaza on the UCLA campus.
The \BS is mounted in front of a building at a height of 8 meters.
During each scene realization, the street level is populated with vehicles and pedestrians whose locations are sampled from a uniform distribution across the scene.
These objects act as dynamic blockage and scattering bodies, while the surrounding buildings and streets act as a static backdrop of the propagation environment.
A random subset of the pedestrians are designated as downlink or uplink users at a height uniformly sampled between $1$ to $1.7$ meters.

Given the geometry of the scene, Sionna computes the signal propagation paths through differentiable ray tracing.
For added richness and realism, we randomly discard 25\% of the output paths to mimic the effects of foliage blockage, small-scale fading, and other phenomena not captured by the ray-tracing simulator.
The remaining propagation paths are summed together to generate narrowband downlink and uplink channels.
This process is repeated multiple times, each with unique vehicle and user placements, in order to populate a dataset of channels that can be used to train and comprehensively evaluate our deep learning model.
To emulate beam management procedures in practical 5G networks~\cite{xue_2024_SurveyBeam}, we assume that the \BS only has knowledge of the most dominant (often \LOS) path to each user, rather than the full channel.
This comprises the user channel information $\mY$ that is input to our deep learning model.

Consistent with \eqref{eq:si-channel}, our model of the self-interference channel $\mH$ consists of a \LOS component and a \NLOS component.
The \NLOS component is produced by Sionna according to the environment's reflective components in each scene realization, and the \LOS component is simulated separately using the near-field spherical-wave model~\cite{jiang_2005_Sphericalwavemodel}.
These two components are then combined according to \eqref{eq:si-channel}, with $\kappa$ determining their relative strengths.
We will use $\kappa = 0$~dB by default, but will sweep its value across a wide range to assess performance.

\begin{table}[!t]
  \centering
  \caption{Model Hyperparameters and Training Details}
  \renewcommand{\arraystretch}{1.1}
  \begin{tabular}{m{.48\linewidth} m{.37\linewidth}}
    \toprule
    \textbf{Parameter} & \textbf{Value} \\
    \midrule
    Embedding dimensions $\Duser$, $\Dsi$       & $256$ \\
    Attention heads                             & $4$ \\
    User encoder layers                         & $2$ \\
    Probing beam synthesizer layers             & $2$ \\
    Serving beam synthesizer layers             & $2$ \\
    Optimizer                                   & AdamW \cite{loshchilov_2019_DecoupledWeight} \\
    Weight decay                                & $0.01$ \\
    Pretraining learning rate (LR)              & $5\times 10^{-5}$ \\
    Pretraining LR schedule                     & Cosine-annealed to $5\times 10^{-6}$ \\
    Fine-tuning LR schedule                     & 1cycle policy with max LR of $3 \times 10^{-4}$~\cite{smith_2019_Superconvergencevery} \\
    \bottomrule
  \end{tabular}
  \label{tab:hyperparameters}
\end{table}

\begin{figure*}[!t]
  \centering
  \subfloat[
    Raw \SSE, $\bar{R}$
  ]{%
    \label{subfig:k-scaling-raw}%
    \includegraphics[width=0.47\linewidth]{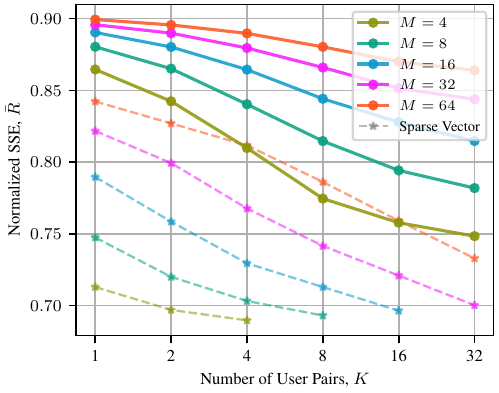}
  }
  \hfil
  \subfloat[
    Effective \SSE, $\bar{R}_{\text{eff}}$
  ]{%
    \label{subfig:k-scaling-eff}%
    \includegraphics[width=0.47\linewidth]{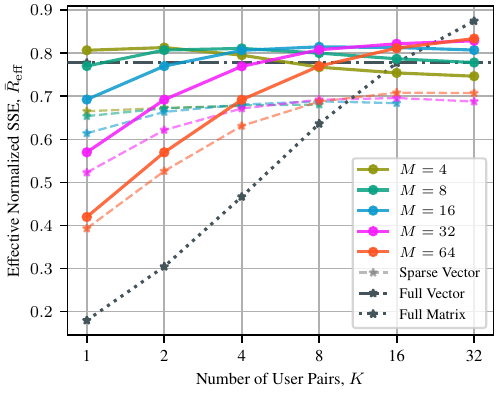}
  }
  \caption{
  Raw and effective normalized \SSE as a function of the coherent user group size $K$ for various probing budgets $M$, with $\kappa=0$~dB. In (a), the Full Matrix and Full Vector baselines attain a raw \SSE of $\bar{R}=1$. In (b), each user is served for $L=56$ time slots.}
  \label{fig:k-scaling}
\end{figure*}

We use a lightweight implementation of the transformer-based architecture detailed in \secref{sec:implement},
which consists of approximately 5.8~million parameters.
The implementation details and training hyperparameters are summarized in \tabref{tab:hyperparameters}.
During each experiment, the model is first pretrained across all combinations of $K$, $M$, and $\kappa$, and then fine-tuned on a specific configuration before evaluating performance.
To standardize training and evaluation, the mean downlink and uplink \gls{snr} upper bounds are normalized to 10~dB, and the mean uplink \INR upper bound is normalized to 40~dB.
To focus exclusively on the effects of self-interference, we assume cross-link interference is negligible, as discussed in \secref{sec:system-model}.

We compare the performance of our proposed approach against three relevant baselines, two of which are upper bounds on performance under differing assumptions:
\begin{itemize}

\item \textbf{\textit{Full Matrix (Upper Bound):}} This baseline assumes the \BS attains the interference-free full-duplex capacity after collecting $\Nt\Nr$ measurements to estimate the entire matrix $\mH$.
Since this attains the capacity, its raw normalized \SSE is $\bar{R}=1$, and its effective \SSE is equal to $\bar{R}_\mathrm{eff} = \frac{KL}{\Nt\Nr + KL}$.
This represents \textit{best-case} performance of any scheme which explicitly estimates $\mH$ from $\Nt\Nr$ measurements.

  \item \textbf{\textit{Full Vector (Upper Bound):}} This baseline assumes the \BS performs \MRT at its transmitter based on perfect knowledge of the users' channels. %
  Then, the \BS is assumed to collect $\Nr$ measurements to estimate the \textit{effective} self-interference channel vector $\vh_k = \mH\vf_k^\star$ for each user pair, where $\vf_k^\star$ is the \MRT transmit beam.
  Afterwards, it is assumed that the \BS attains the interference-free full-duplex capacity, meaning its raw normalized \SSE is $\bar{R}=1$, and its effective \SSE is equal to $\bar{R}_\mathrm{eff} = \frac{KL}{K\Nr + KL}$.
  This represents \textit{best-case} performance of any scheme which \textit{explicitly} estimates $\vh_k$ for each user pair from $\Nr$ measurements.
    
    \item \textbf{\textit{Sparse Vector:}}
    Using the same definition of $\vh_k$ as in the \textit{Full Vector} baseline (but with partial knowledge for \MRT), this baseline assumes the \BS measures $\vh_k$ for each user pair by sweeping  $M/K \geq 1$ probing beams at its receiver; this means a total of $M$ probing measurements are collected across user pairs.
    Provided with the same knowledge as our proposed scheme, these probing beams are designed based on the user channel knowledge to minimize the posterior error variance of $\vh_k$ using site-specific statistics computed from training data.
    After collecting all $M$ probing measurements, the \BS computes the LMMSE estimate of the effective self-interference channel vector $\vh_k$ for each user along with its posterior error covariance $\mC_{k}$, sharing probing measurements across user pairs in the process.
    For each user pair, the \SINR-maximizing receive beam $\vw_k^\star$ is then designed in an LMMSE fashion using the channel estimate $\hat\vh_k$ and the error covariance $\mC_{k}$.
    We refer to this baseline as \textit{Sparse Vector} since it only partially measures the effective self-interference channel vectors $\vh_k$.%
    \footnote{We extensively evaluated several other baseline schemes under the same measurement budgets, including compressive sensing, low-rank matrix recovery, full matrix probing followed by alternating optimization beam design, and a scheme where probing beams were fixed while keeping serving beam synthesis learnable. 
    Within this setting, the self-interference channels in our dataset were not strictly sparse, preventing compressive estimators from attaining the low estimation error needed to cancel strong self-interference.
    The \textit{Sparse Vector} approach proved to be the most competitive baseline, so the others have been omitted for conciseness.}
\end{itemize}

\section{Performance Evaluation} \label{sec:results}

In the results that follow, we assess performance using the raw normalized \SSE
\begin{align}
  \bar{R} = \frac{1}{K} \sum_{k=1}^K \bar{R}_k(\vf^\star_k, \vw^\star_k),
\end{align}
along with the \textit{effective} normalized \SSE
\begin{align}
  \bar{R}_\mathrm{eff} = \frac{KL}{M+KL} \, \bar{R},
\end{align}
which accounts for the measurement overhead associated with acquiring self-interference channel knowledge.
Other practical overheads, including user beam alignment, beam switching, and control signaling, are omitted here since they affect all of the evaluated schemes equally.
Note that, in both, the spectral efficiency has been normalized to the true full-duplex capacity, based on perfect user channel information, rather than the partial knowledge provided during initial beam alignment.

\subsection{Coherent User Group Size, $K$}

To begin our performance evaluation, we first investigate the effect of the coherent group size $K$; recall, $K$ represents the number of user pairs over which the self-interference channel $\mH$ remains static and can thus be thought of as a proxy for its coherence time.
In \figref{fig:k-scaling}, we fix $\kappa=0$~dB and plot both the raw and effective normalized \SSE as a function of $K$ for various numbers of probing measurements $M$.
We also plot performance of the three baselines.
Since $\Nt=\Nr=16$ in this case, the \textit{Full Vector} baseline requires $K\Nr=16K$ measurements and the \textit{Full Matrix} baseline requires $\Nt\Nr=256$ measurements; both attain a raw normalized \SSE of $1$.

In \figref{subfig:k-scaling-raw}, we see that our proposed scheme outperforms the \textit{Sparse Vector} baseline (dashed lines) in terms of raw \SSE across the board.
For instance, when $K=4$, our scheme with $M=4$ probing measurements achieves the same performance as \textit{Sparse Vector} with $M=64$ measurements; this can be attributed to both more informative user-specific probing and superior serving beam synthesis.
When $M$ is fixed and $K$ is increased, our scheme's performance naturally degrades, since the collected measurements are forced to generalize across more user pairs.
The measurements thus become less informative for any individual user pair, degrading the model's ability to reliably cancel self-interference.
Fixing $K$ and increasing the number of measurements $M$, conversely, provides more informative implicit channel knowledge and in turn yields an increase in raw \SSE.

\begin{figure}[t]
    \centering
    \includegraphics[width=\linewidth]{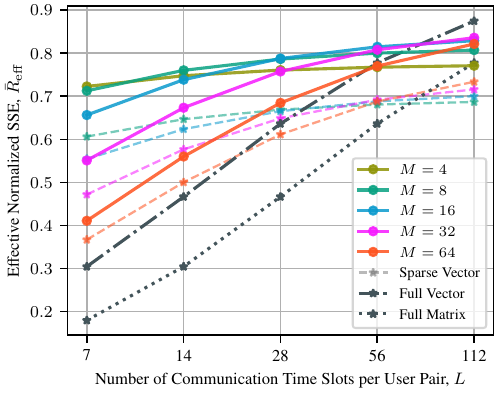}
    \caption{Effective normalized \SSE as a function of the number of communication time slots per user $L$ for various probing budgets $M$, with $\kappa=0$~dB and $K=8$ user pairs.}
    \label{fig:l-scaling-eff}
\end{figure}

To account for the overhead associated with more measurements, \figref{subfig:k-scaling-eff} repeats this evaluation in terms of the \textit{effective} \SSE with each user pair served for $L=56$ time slots%
\footnote{In 5G NR, a single slot is 14 OFDM symbols \cite{3gpp38211}, and with analog beamforming, each probing measurement occupies an entire OFDM symbol. If each user is served for 2--8 slots, then this corresponds to $L=28$ to $L=112$ symbols. Considering operation at 28~GHz (FR2) with a subcarrier spacing of 120~kHz, then $L=56$ symbols corresponds to around $500$ \textmu{}sec. If the coherence time of $\mH$ is around 4 ms, for instance, then $K=8$ user pairs could be served from a single set of probing measurements $\vz$.}, where a more nuanced trend emerges.
It can be seen that increasing $M$ can only be justified when $K$ also increases, i.e., when the coherence time of $\mH$ becomes longer, since this amortizes the measurement overhead across multiple user pairs.
Our proposed approach outperforms all baselines for all $K$, except for \textit{Full Matrix} when $K=32$, corresponding to a long coherence time. %
This superiority indicates that our proposed scheme exhibits a more favorable trade-off between measurement overhead and raw \SSE than the \textit{Sparse Vector} baseline.
In other words, for a given number of measurements $M$, our scheme tends to offer a higher raw \SSE, thus yielding a higher effective \SSE.

\begin{figure}[!t]
    \centering
    \includegraphics[width=\linewidth]{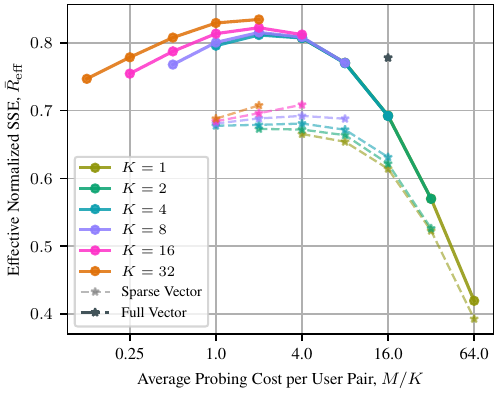}%
    \caption{Effective normalized \SSE as a function of the average probing cost per user pair $M/K$ for various coherence group sizes $K$. Each user is served for $L=56$ time slots.}
    \label{fig:mpk-scaling-eff}
\end{figure}

\begin{figure*}[!t]
    \centering
    \subfloat[
    $K=1$ user pair
    ]{
        \label{subfig:patterns-k1}
        \includegraphics[width=0.47\linewidth]{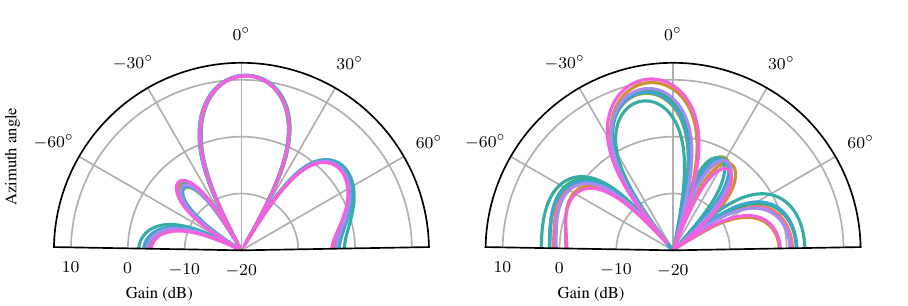}
    }
    \hfil
    \subfloat[
    $K=32$ user pairs
    ]{
        \label{subfig:patterns-k32}
        \includegraphics[width=0.47\linewidth]{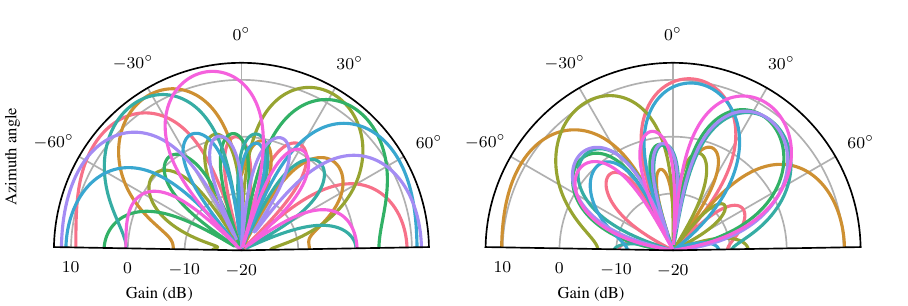}
    }
    \caption{
        The beam patterns (azimuth cuts) of $M=8$ synthesized probing beam pairs when serving (a) $K=1$ user pair and (b) $K=32$ user pairs, illustrating how the model prioritizes noise suppression when $M/K$ is high and subspace exploration when $M/K$ is low.
    }
    \label{fig:patterns}
\end{figure*}

\begin{figure*}[!t]
    \centering
    \subfloat[
    Raw \SSE, $\bar{R}$
    ]{
        \label{subfig:kappa-scaling-raw}
        \includegraphics[width=0.47\linewidth]{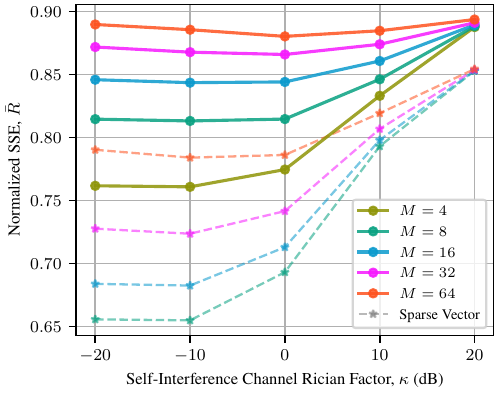}
    }
    \hfil
    \subfloat[
    Effective \SSE, $\Rne$
    ]{
        \label{subfig:kappa-scaling-eff}
        \includegraphics[width=0.47\linewidth]{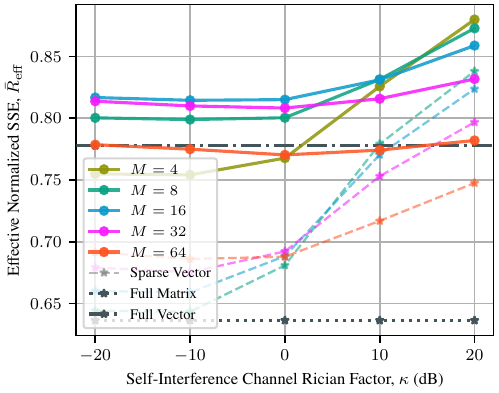}
    }
    \caption{
        Raw and effective normalized \SSE as a function of the \SI channel Rician factor $\kappa$ for various probing budgets $M$, with $K=8$ user pairs. In (b), each user is served for $L=56$ time slots.
    }
    \label{fig:kappa-scaling}
\end{figure*}

\subsection{Communication Time Slots, $L$}
The net cost of measurement overhead, captured by the fraction $\frac{KL}{M+KL}$, depends on the number of time slots $L$ each user pair is served.
This motivates our next evaluation in \figref{fig:l-scaling-eff}, which shows the effective \SSE as a function of $L$ for $K=8$ user pairs.
With smaller $L$, the measurement overhead is magnified, and the highest effective \SSE is achieved with fewer measurements.
For this reason, the \textit{Full Matrix} and \textit{Full Vector} baselines exhibit poor performance when $L$ is small, given they require $256$ measurements in total and $16$ measurements per user, respectively.
The \textit{Sparse Vector} baseline outperforms these upper bounds below $L=56$ thanks to its lower overhead. 
Our scheme outperforms all baselines for all $L$, except for \textit{Full Vector} when $L=112$.
Taken together with \figref{subfig:k-scaling-eff}, these results suggest that optimal deployment of our proposed solution involves choosing $M$ and $K$ based on the coherence time of the self-interference channel relative to the duration that users are served.

\subsection{Average Probing Cost, $M/K$}

Our approach collects $M$ measurements at once and reuses the information contained in these measurements across all $K$ user pairs.
It is interesting to ask whether this is superior to an approach where the $M$ measurements are divided up across smaller groups of users, allowing each subset of measurements to focus on fewer user pairs.
To investigate this, \figref{fig:mpk-scaling-eff} plots the effective \SSE as a function of $M/K$, the average probing cost per user pair.

Let us first consider the case of $M/K=1$, meaning the number of measurements $M$ equals the total number of user pairs $K$.
If $M=K=4$, for example, then $4$ user pairs would be grouped together and $4$ measurements would be spent per user group, for an effective cost of one probing measurement per user pair.
If $M=K=64$, this would yield the same per-user-pair probing cost, but would be accomplished with groupings of $64$ user pairs and spending $64$ measurements per group.
While these two groupings exhibit the same relative overhead, \figref{fig:mpk-scaling-eff} indicates that a larger grouping results in superior performance, both by our proposed scheme as well as the \textit{Sparse Vector} baseline.
This suggests that the measurements in $\vz$ are mutually informative across user pairs, resulting in larger groups benefiting more from shared probing.
This is because certain portions of the self-interference channel $\mH$ may be relevant to multiple user pairs within a coherent user group due to correlations in their channels.
This is most pronounced when $M/K$ is low, since this is where performance is constrained by limited self-interference channel knowledge.
These results suggest that it is best to group as many user pairs as possible and perform a single sweep of $M$ probing measurements, as permitted by the coherence time of $\mH$.

\begin{figure*}[!t]
  \centering
  \includegraphics[width=\linewidth]{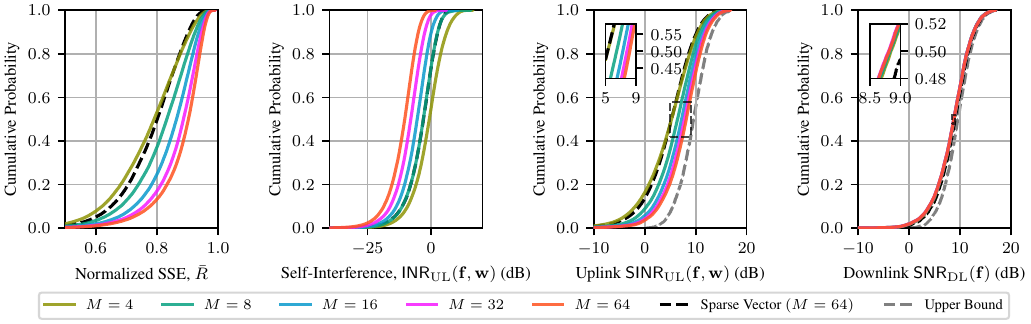}
  \caption{CDFs of the normalized \SSE, self-interference, uplink \SINR, and downlink \SNR for various probing budgets $M$, with $K=8$ user pairs and $\kappa=0$~dB.}
  \label{fig:cdf-K8}
\end{figure*}

\subsection{Learned Probing Beam Patterns}

To gain some intuition on how our model synthesizes its probing beams, \figref{fig:patterns} depicts sample beam patterns for $M=8$ probing beams when $K=1$ (a single user pair) and $K=32$ (a large user group), representing two contrasting regimes to visualize the model's behavior. 
When $K=1$, the model repeatedly probes along the channel of the user to suppress noise, resulting in highly correlated probing beams. 
In contrast, with $K=32$ user pairs, the model synthesizes uncorrelated probing beams to explore the broader interference subspace seen across all $32$ user pairs. 
This highlights the model's ability to adaptively balance noise suppression against subspace exploration based on the number of user pairs being served $K$ and the measurement budget $M$.

\begin{figure*}[!t]
  \centering
  \subfloat[
    Raw \SSE, $\bar{R}$
  ]{%
    \label{subfig:nr-scaling-raw}
    \includegraphics[width=0.47\linewidth]{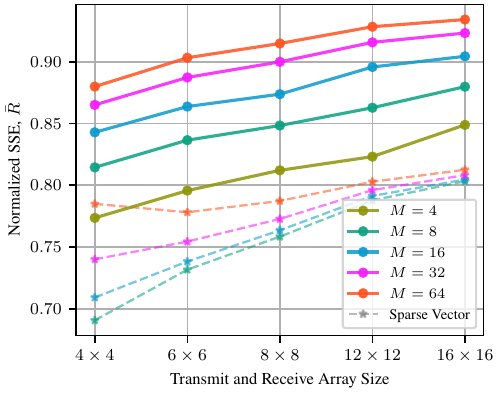}
  }
  \hfil
  \subfloat[
    Effective \SSE, $\Rne$
  ]{%
    \label{subfig:nr-scaling-eff}
    \includegraphics[width=0.47\linewidth]{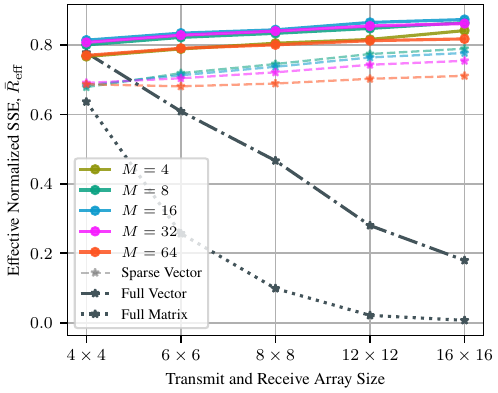}
  }
  \caption{
    Raw and effective normalized \SSE as a function of the transmit and receive array size for various probing budgets $M$, with $K=8$ user pairs and $\kappa=0$~dB. In (b), each user is served for $L=56$ time slots.
  }
  \label{fig:nr-scaling}
\end{figure*}

\subsection{Structure of the Self-Interference Channel, $\kappa$}
\label{subsec:kappa-scaling}

We now assess performance as a function of the self-interference channel structure by varying its Rician factor $\kappa$ from $-20$~dB to $20$~dB in \figref{fig:kappa-scaling}.
We observe that, as $\kappa$ increases, the channel becomes more \LOS-dominant and more deterministic, allowing our proposed scheme to perform well with fewer probing measurements.
Even for very low $\kappa$, where the channel exhibits greater stochasticity, the proposed scheme remains quite competitive.
A similar trend can also be seen in the \textit{Sparse Vector} baseline.

Accounting for measurement overhead, net performance is seen in the effective \SSE of \figref{subfig:kappa-scaling-eff}.
At low $\kappa$, a modest number of measurements is preferable, e.g., $M=16$, since they provide greater value when the channel is more stochastic.
As $\kappa$ grows, the raw performance gain of additional probing diminishes, since the channel becomes more deterministic, allowing performance with only a few measurements to climb and eventually overtake the case of heavy probing.
Note that our proposed scheme and \textit{Sparse Vector} both perform relatively poorly with $M=64$; this is because the marginal gain in raw \SSE does not justify the high overhead, in this case.

\begin{figure*}[!t]
    \centering
    \includegraphics[width=\linewidth]{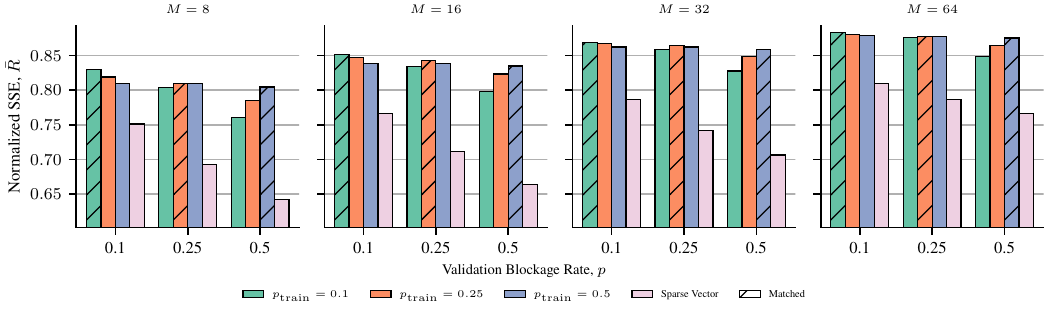}
    \caption{Raw normalized \SSE under blockage mismatch across various probing budgets $M$ when evaluating models trained under nominal path blockage rates $p_\mathrm{train} \in \{10\%, 25\%, 50\%\}$ and tested across mismatched blockage rates $p$, with $K=8$ user pairs, $\kappa=0$~dB, and $\Nt=\Nr=16$.}
    \label{fig:mismatch-blockage}
\end{figure*}

\subsection{Performance Distribution}

The CDFs in \figref{fig:cdf-K8} provide a more detailed view of several relevant metrics in order to further explore the performance distribution.
As established before, increasing the number of measurements $M$ reliably increases the raw \SSE, and this is thanks to an improved ability to cancel self-interference.
This results in an increase in uplink \SINR exceeding that of \textit{Sparse Vector}, due to our model's more informative channel knowledge and its minor sacrifices in downlink \SNR. 
This further confirms that our proposed scheme more optimally trades off downlink and uplink quality to maximize \SSE (a non-convex problem), compared to the \textit{Sparse Vector} baseline.
The upper bounds on uplink \SINR and downlink \SNR shown in \figref{fig:cdf-K8} also provide insight into the fairness of our scheme.
Across user pairs, both metrics closely approach their respective upper bounds, especially when compared to the \textit{Sparse Vector} baseline. 
This confirms that, although fairness between the uplink and downlink is not explicitly enforced in our problem formulation, the learned policies operate in a fairly balanced manner, delivering high beamforming gain to both users while effectively suppressing self-interference on the uplink.

\subsection{Antenna Array Size, $\Nt$ and $\Nr$}
Next, \figref{fig:nr-scaling} examines how performance scales with the size of the \BS's transmit and receive antenna arrays, which we assume are identical in size; while changing the arrays' size, we keep all other parameters constant, including the total transmit power.
We begin with $4\!\times\!4$ \UPAs at the transmitter and receiver and then scale up to $16\times16$ \UPAs.
This corresponds to self-interference channel matrices that range from $16^2=256$ entries to $256^2=65,\!536$ entries, which illustrates the high measurement overhead associated with the \textit{Full Matrix} baseline.

The raw \SSE shown in \figref{subfig:nr-scaling-raw} improves with the array size for the proposed scheme and the baselines, reflecting the beamforming gains afforded by larger arrays---both in terms of \SNR gain and interference reduction.
Once the probing overhead is accounted for in \figref{subfig:nr-scaling-eff}, however, our proposed scheme and \textit{Sparse Vector} diverge sharply from the \textit{Full Matrix} and \textit{Full Vector} baselines.
The effective \SSE offered by these latter two baselines plummets as the antenna count increases, since their overhead scales either linearly or quadratically with $\Nt=\Nr$.
Performance with our proposed approach and \textit{Sparse Vector} grows steadily, consistent with the behavior in \figref{subfig:nr-scaling-raw}, further demonstrating our model's ability to efficiently gather and harness \textit{implicit} channel knowledge, even as the channel grows in size.
This can be explained by the simple fact that the physical sources of \NLOS self-interference (e.g., prominent reflectors, blockage) are not affected by the number of antennas, meaning the required number of probing measurements should also not be impacted.
It is also worth noting that this increase in antenna count does not come at the cost of substantial increases in computational complexity, since our model's transformer-based construction is based on a fixed embedding size, as mentioned in \secref{sec:implement}.
Overall, these results demonstrate that our proposed approach scales gracefully to large antenna arrays, making it an attractive solution for the coming 6G era as networks pursue \BSs with an ever-increasing number of antennas.

\subsection{End-to-End Inference Latency}

\begin{figure}[!t]
    \centering%
    \includegraphics[width=\linewidth,height=\textheight,keepaspectratio]{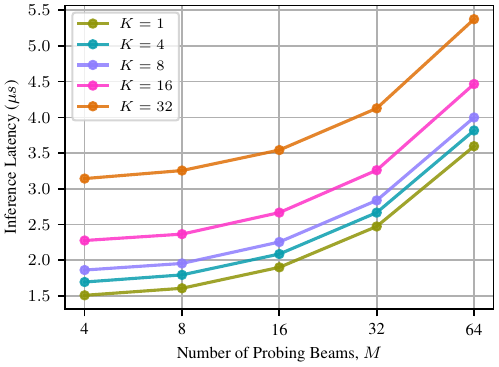}%
    \caption{End-to-end inference latency as a function of $M$ across $K \in \{1, 4, 8, 16, 32\}$, measured on an NVIDIA RTX 5090 using PyTorch with \texttt{torch.compile} in \texttt{bfloat16} precision with $\Nt=\Nr=16$.}%
    \label{fig:latency}%
\end{figure}

In addition to probing overhead, runtime computational latency is also a concern in practical deployments.
To assess this, \figref{fig:latency} shows the end-to-end inference latency of our proposed model as a function of the measurement budget $M$ across various user group sizes $K$.
Across all evaluated regimes, the inference latency remains remarkably low, requiring only about $2.25$~\textmu{}s for typical operating conditions and staying below $5.5$~\textmu{}s under denser user groupings and large measurement budgets.
Furthermore, because array dimensions only affect the lightweight input and output linear projections, our model experiences negligible latency scaling with the antenna counts $\Nt$ and $\Nr$.
This low computational burden suggests that our approach is viable in real-world deployments, executing beam synthesis in real time without introducing noticeable scheduling delay.

\begin{figure}[!t]
    \centering%
    \includegraphics[width=\linewidth,height=\textheight,keepaspectratio]{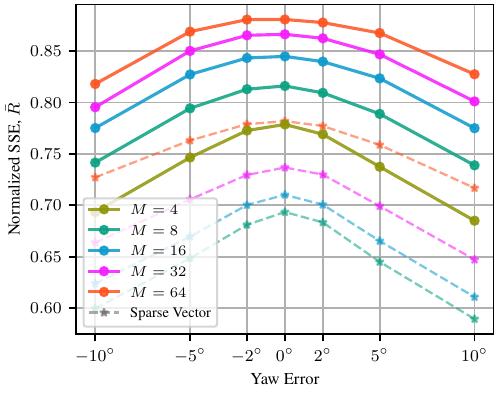}%
    \caption{Raw normalized \SSE as a function of base station array mechanical yaw error for various probing budgets $M$, with $K=8$ user pairs, $\kappa=0$~dB, and $\Nt=\Nr=16$.}%
    \label{fig:mismatch-yaw}%
\end{figure}

\subsection{Robustness to Mismatch: Blockage and Array Positioning}
Finally, to assess the robustness of our proposed approach against mismatch, we evaluate model performance when deployment conditions differ from those seen during training.
We specifically consider two cases: blockage mismatch and array positioning mismatch. 
Thus far, all of the results presented have been with our model trained and tested on a fixed $25\%$ blockage rate.
In real-world settings, however, moving obstacles and changing foliage introduce varying levels of blockage.
We model this by training under nominal blockage rates $p_\mathrm{train} \in \{10\%, 25\%, 50\%\}$ and testing under mismatched rates $p$ across different probing budgets $M$.
As shown in \figref{fig:mismatch-blockage}, the spectral efficiency drops by only a few percent under blockage mismatch, and our scheme consistently outperforms the \textit{Sparse Vector} baseline across the board, whose performance degrades further under higher blockage rates due to increased stochasticity. 
This suggests that our model is robust to changes in the particular blockage level of its environment.

Another practical source of mismatch is in mounting tolerances and structural vibrations that can cause the physical orientation of the base station to deviate from what was assumed during training or to drift over time.
We assess the effects of this by evaluating a model trained on the nominal array placement under varying yaw misalignments (rotation about its vertical axis) over a range of $\pm 10^\circ$ for $K=8$ user pairs across different probing budgets $M$.
As shown in \figref{fig:mismatch-yaw}, the normalized sum spectral efficiency degrades gracefully across these angular perturbations, and the proposed scheme continues to maintain its advantage over the \textit{Sparse Vector} baseline.
Together, these results confirm that the model's performance remains resilient to blockage variations and physical misalignment, even though it was not explicitly trained to handle such mismatch.

\section{Conclusion} \label{sec:conclusion}

This work introduces a novel approach to designing the transmit and receive beams of a full-duplex massive \MIMO \BS that avoids the prohibitively high overhead associated with explicit estimation of the self-interference channel $\mH$.
This is accomplished through site-specific training of a transformer-based deep learning model that learns to efficiently probe the portions of $\mH$ most relevant to the particular users being served.
By conditioning this probing on partial user channel knowledge, the model focuses only on the channel features necessary to synthesize transmit and receive beams that deliver high gain and suppress self-interference.

Simulation results using ray-tracing demonstrate that our proposed scheme consistently outperforms both conventional and compressive baselines across various user group sizes and probing budgets, achieving superior spectral efficiency with fewer measurements.
Furthermore, thanks to the added spatial degrees of freedom, the performance of our approach actually improves as the base station scales to larger antenna arrays, at the expense of only modest increases in computational complexity.

Valuable future directions include extending the proposed framework to wideband, frequency-selective self-interference channels and developing channel tracking/prediction to exploit temporal channel correlations.
More broadly, this framework could also be extended to \gls{isac} applications, where sparse, site-specific probing can simultaneously detect targets and suppress self-interference and environmental clutter.

\section*{Acknowledgments}

This work used computational and storage services associated with UCLA's Hoffman2 Cluster, operated by the UCLA Office of Advanced Research Computing.

\bibliographystyle{IEEEtran}
\bibliography{refs}

\end{document}